\documentclass{mem}
\usepackage{natbib}\usepackage{txfonts}\usepackage{balance}
\usepackage{graphicx}
\usepackage{txfonts}
\usepackage[a4paper]{hyperref}
\idline{79}{1}
\begin{document}
\def\teff{$T\rm_{eff }$}
\def\kms{$\mathrm {km s}^{-1}$}

\title{
Lithium-rich giants in the\\ Sagittarius dSph Tidal Streams
}

   \subtitle{}

\author{
L. \,Monaco\inst{1} 
\and P. \,Bonifacio\inst{2,3}
          }

  \offprints{L. Monaco}

\institute{
ESO - European Southern Observatory
Alonso de Cordova 3107
Casilla 19001, Santiago 19, Chile
\and
Istituto Nazionale di Astrofisica --
Osservatorio Astronomico di Trieste, Via Tiepolo 11,
I-34131 Trieste, Italy
\and
CIFIST, Marie Curie Excellence Team and GEPI,
Observatoire de Paris, 5, place Jules Janssen, 92195 Meudon, France
\email{lmonaco@eso.org}
}

\authorrunning{Monaco \& Bonifacio}

\titlerunning{Lithium-rich stars in the Sgr Streams}

\abstract{

The nature and origin of Lithium rich giant stars is still matter of debate.  In
the contest of our spectroscopic survey of giants in the tidal streams of the 
Sagittarius dwarf spheroidal (dSph) galaxy,  we present the serendipitous
discovery of 2 super Li-rich stars (A(Li)$>$3.5-4.0).  Besides D461 in Draco,
these are the only Li-rich stars known in a Local Group dSph galaxy. The high Li
abundance and the low mass of these stars support their origin as due to fresh
Li production in the stars associated with some kind of extra-mixing process.

}
\maketitle{}

\section{Introduction}

Since the serendipitous discovery by \citet{mckellar} that the giant star WZ Cas
was anomalously rich in Lithium \citep[see also][]{mckellar41}, the nature and
origin of such Li-rich stars has been  object of considerable interest. These
stars are very rare, which can  immediately  be connected to the fact that the
Li-rich status is experienced during a short-lived phase. Up to 1991 only 8 such
stars were known \citep[see][for a brief account of the  literature prior to
this date]{faraggiana}. Currently this number has risen to over 50, thanks to
dedicated surveys \citep{mc}. A few Li-rich giants are known in globular and
open clusters \citep{kraft,hill} and  one such star has also been found in the
Local Group dwarf spheroidal Draco \citep{imma}.

As a star evolves off the Main Sequence up the Red Giant Branch (RGB), the
convective envelope brings to the surface material which experienced
temperatures in excess of $2\times 10^{6}$K and was, therefore, depleted in
Lithium. This mixing of unprocessed material with Li-depleted material is
referred to as dilution.  The dilution factor  has been estimated to be 1.8 dex
for a 3~M$_\odot$~ star and 1.5 dex for a 1~M$_\odot$ ~ star
\citep{ibena,ibenb}. Thus, assuming a star begins its life with 
A(Li)\footnote{A(Li)=log(Li/H)+12}=3.0, any giant  with A(Li)$>1.5$ should be
considered as ``Li rich'', i.e. with a Li content above what expected by
standard models. Typically 1-2\% of giant stars is Li-rich \citep{cb00} and 
just a handful of stars have Li abundances exceeding the  meteoritic value. 

The proposed explanations for the Li-rich phenomenon fall
into three categories:
\begin{enumerate}
\item preservation:  mixing is inhibited or less efficient and 
the dilution is lower than expected from standard models predictions;
\item planet or brown dwarf engulfment: the star has swallowed
one or more planets or brown dwarfs, which are rich in Li,
since none or little has been destroyed during their formation;
\item Li formation: fresh Li is produced through the Cameron-Fowler mechanism 
\citep{cf71}. 
\end{enumerate}

A less efficient dilution process should preserve $^9$Be and the
$^{12}$C/$^{13}$C ratio as well. On the contrary, $^9$Be is strongly depleted in
the sample observed by \citet{castilho} and \citet{melo}. Li-rich giants  also
present low  $^{12}$C/$^{13}$C ratios \citep[see][and references therein]{bala}.
Therefore, mixing seems at play in these stars.

On the other hand, an  engulfment episode should enrich the star also in $^9$Be,
$^6$Li and $^{11}$B. We already noticed the low $^9$Be abundances observed in
Li-rich giants. Apparently, Li-rich giants are also devoided of $^6$Li
\citep[see][and references therein]{bala,drake}.

Furthermore, the first two scenarios can hardly account for  the existence of
giants with Li abundances exceeding the meteoritic value \citep{siess,bala}.
Therefore, Li production appears as the most likely cause for the observed Li
abundances, although  probably not all of the Li-rich stars share a common
origin \citep{drake}.

In the contest of our high resolution spectroscopic survey of stars in the tidal
streams of the Sagittarius dwarf spheroidal galaxy (Sgr dSph), we present here
the serendipitous discovery of two Li-rich giants.

\section{Li-rich stars in the Sgr Streams}

The Sagittarius dwarf spheroidal galaxy \citep{igi94} is currently disrupting
into the Milky Way. It presents a very significant core remnant  (30$^\circ$
tidal radius), and its giant tidal streams indicate that the disruption process
is still ongoing.  Recently, \citet[][hereafter M03]{maje03} traced the Sgr
tidal streams all over the sky using 2MASS data. Using different high resolution
facilities, we observed  a sample of 2MASS selected giants belonging to the Sgr
streams \citep[][hereafter M07]{monaco07}. 

\begin{figure}[]
\resizebox{\hsize}{!}{\includegraphics[clip=true]{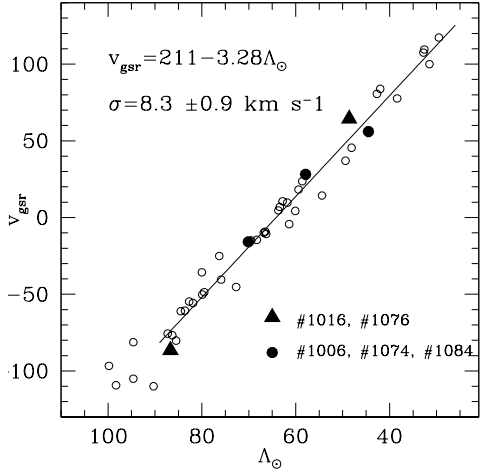}}
\caption{
\footnotesize
Measured radial velocities as a function of the
longitude of the Sgr orbital plane for the Sgr stream stars observed with UVES.
Filled symbols mark stars presenting a clear Li resonance line.
}
\label{li_vhel}
\end{figure}

In particular, 46 stars belonging to the Sgr southern stream were observed with
the UVES spectrograph mounted at the VLT (Paranal, Chile).  Details of the
observations are reported in M07. In Fig.~\ref{li_vhel} we present for these
stars the measured radial velocity (in the galactic standard of rest, v$_{gsr}$)
as a function of the Sgr longitude scale ($\Lambda_\odot$) along the orbital
plane (see M03 for definitions and details). As can be seen, the Sgr southern
stream is a very coherent and dynamically cold structure 
\citep[$\sigma$=8.3$\pm$0.9\kms, see also][]{maje04}. Therefore, we are quite
confident that the observed stars were once part of the Sgr galaxy.

\begin{figure}[]
\resizebox{\hsize}{!}{\includegraphics[clip=true]{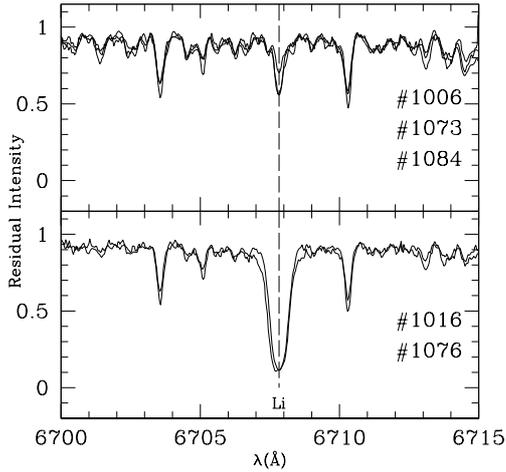}}
\caption{
\footnotesize
Sample of the UVES spectra of the 5 stream stars presenting a clear Li 
resonance line.
}
\label{li}
\end{figure}

\begin{figure}[]
\resizebox{\hsize}{!}{\includegraphics[clip=true]{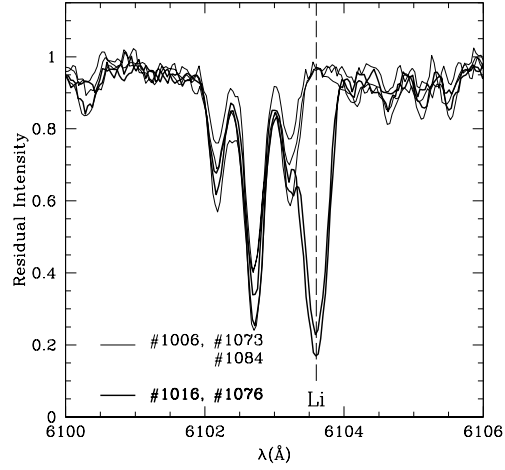}}
\caption{
\footnotesize
Same as Fig.~\ref{li} but in the spectral region around the 6103.6$\AA$ Li 
subordinate line. Only \#1016 and \#1076 present a strong Li line.
}
\label{lisub}
\end{figure}

The five stars marked with solid symbols\footnote{Identification numbers refer
to Table~1 in M07.} in Fig.~\ref{li_vhel} present a clearly detectable Lithium
resonance line (see Fig.~\ref{li}). Stars \#1016 and \#1076, in particular, 
present a very strong absorption line. These two stars also show a strong
Li-subordinate line at 6103.6$\AA$ (Fig.~\ref{lisub}), while this line is
completely absent in the remaining three stars.

\section{Chemical Abundances}

We performed a preliminary chemical abundance analysis for stars \#1016 and
\#1076. The adopted atmospheric parameters are reported in Table~\ref{atmpar}.
Effective temperatures were derived from the 2MASS dereddened (J-K) infrared
color, adopting the \citet{alonso} calibrating relations. Putative distance for
the program stars were derived assuming that Sgr stream stars follow the same
color-magnitude relation as stars in the core of Sgr (see M07). After correcting
for their distance and reddening, gravity was derived by comparison with
theoretical isochrones (see M07 and Fig.\ref{cmd}). ATLAS model atmospheres 
were calculated using the atmospheric parameters reported in Table~\ref{atmpar}
and the Opacity Distribution  Functions of \citet{newodf} under the Local
Thermodynamic Equilibrium (LTE) approximation. We measured equivalent widths
(EWs) on the spectra for a selected sample of Fe~I lines using the standard IRAF
task {\it splot}. Iron abundances were derived from the measured EWs using the
calculated model atmospheres within the WIDTH code.  The GNU-Linux ported
version \citep{sbordone04} of both the WIDTH and ATLAS codes \citep{k93} were
employed.   Microturbulent velocities ($\xi$) for each star were determined
minimizing the dependence of the iron abundance from the EW. 

\begin{table}
\caption{Atmospheric parameters adopted for stars \#1016 and \#1076}
\label{atmpar}
\begin{center}
\begin{tabular}{lccc}
\hline
Star \# & T$_{eff}$ & log g & $\xi$\\
\hline
1016 & 3800 & 0.8 & 1.9\\
1076 & 3750 & 0.7 & 2.0\\ 
\hline
\end{tabular}
\end{center}
\end{table}

Oxygen and Carbon abundances were obtained performing spectral synthesis  of the
6300$\AA$ line and around the G-band spectral region (4300-4340\AA),
respectively. Synthetic spectra were calculated adopting the ATLAS models
described above and  using the SINTHE code \citep{k93,sbordone04}.  The final
abundances were derived taking into account the coupling between C and O
abundances \citep[see][]{gratton}.

Finally, Li abundances were derived by spectral synthesis of the resonance and 
subordinate lines, considering also the calculated values for the C and O
abundances.

Table~\ref{abun} reports the measured Fe, C, O and Li abundances.

\begin{table*}
\caption{Chemical abundances measured in \#1016 and \#1076}
\label{abun}
\begin{center}
\begin{tabular}{lccccccc}
\hline
Star \# & [Fe/H] & A(Li)$_{@6707.8\AA}$ &  A(Li)$_{@6103.3\AA}$ & C/O  & [C/O] & [C/Fe] & [O/Fe]\\
\hline
1016  &$ -0.78\pm0.23 $ & $4.29\pm0.01 $& $4.20\pm0.03 $ &$+0.29$ & $-0.22$ & $+0.09 $ & $+0.31$ \\
1076  &$ -0.74\pm0.27 $ & $3.58\pm0.02 $& $3.50\pm0.03 $ &$+0.15$ & $-0.50$ & $-0.37 $ & $+0.13$ 
\\
\hline
\end{tabular}
\end{center}
\end{table*}

\section{Discussion}

The abundances reported in Table~\ref{abun} confirm that both \#1016 and \#1076
are Li-rich stars. Actually, the Li abundances derived from both the resonance
and the subordinate lines are higher than the meteoritic value. Besides these
two stars, only one Li-rich star is known in a Local Group dSph galaxy: the
Carbon star D461 in Draco \citep[C/O=3-5,][]{imma}. On the other hand, \#1016
and \#1076 are O-rich stars, given the low measured C/O ratio (0.15-0.29).  We
also note that the measured [Fe/H] abundances are perfectly in line with the
mean stream abundance measured by M07.

As already noted in \S1, preservation of primordial Li or planet/brown dwarf
engulfment can hardly account for the existence of super Li-rich giants
\citep{bala}.  Therefore, the super meteoritic Li abundance measured in \#1016
and \#1076 points toward a production of fresh Li in these stars. 

We point out that the abundances reported in Table~\ref{abun}  were calculated
under the LTE approximation.  However, non-LTE approaches usually results in
even higher Li abundances \citep{nonlte1,nonlte2}.

Li-production can happen through the Cameron and Fowler mechanism \citep{cf71}.
According to this model, $^3$He in the convective envelope is converted to
$^7$Be. Then, $^7$Be must be circulated to the star surface where it decays into
$^7$Li. 

\begin{figure}[]
\resizebox{\hsize}{!}{\includegraphics[clip=true]{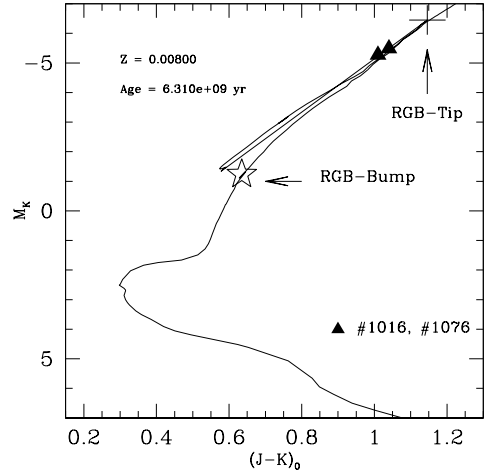}}
\caption{
\footnotesize
The positions of stars \#1016 and \#1076 are marked in the infrared absolute
color-magnitude diagram (triangles). Distance and  reddening were adopted from
\citet{maje04}. A isochrone matching the position of the two stars is plotted.
The RGB-tip and RGB-bump positions along the isochrone are also labelled.
}
\label{cmd}
\end{figure}

As can be seen from Fig.~\ref{cmd}, \#1016, \#1076 are low mass star
(M=1-1.5M$_\odot$) on the  Red/Asymptotic Giant Branch close to the RGB tip.
The adopted isochrone  \citep[][]{leo} has age and metallicity compatible with
current estimates of these parameters for the main Sgr stellar population
\citep[see][]{monaco05,michele06}.

In such low mass stars, temperatures high enough for $^3$He burning are reached
in the vicinity of the hydrogen burning shell. Therefore, an extra-mixing
process is required to circulate material from the convective envelope in and
out of this region. 

In summary, the high Li-abundance measured in our stars and their low masses  
suggest the production of fresh Li should have been coupled with some kind of
rather extreme extra-mixing process in \#1016 and \#1076 \citep[see
also][]{cbp,utten}.


\bibliographystyle{aa}

\end{document}